\documentclass[showpacs,preprintnumbers,amsmath,amssymb]{revtex4}

\usepackage{graphicx}
\usepackage{dcolumn}
\usepackage{bm}
 
\def\mbi#1{\mbox{\bfseries\itshape #1}} 

\begin{document}

\preprint{APS/123-QED}

\title{Constraints on the neutrino mass and the primordial magnetic field from the matter density
fluctuation parameter $\sigma_8$}

\author{Dai G. Yamazaki$^{1}$}
 \email{yamazaki@asiaa.sinica.edu.tw}
\author{Kiyotomo Ichiki$^{2}$}%
\author{Toshitaka Kajino$^{3,4}$}%
\author{Grant. J. Mathews$^{5}$}%
\affiliation{%
$^{1}$Institute of Astronomy and Astrophysics, Academia Sinica,
11F of Astronomy-Mathematics Building,
National Taiwan University.No.1, Roosevelt Rd, Sec. 4 Taipei 10617, Taiwan, Republic of China
}%
\affiliation{%
$^{2}$Department of Physics and Astrophysics, Nagoya Univresity, Nagoya 464-8602, Japan
}%
\affiliation{%
$^{3}$
National Astronomical Observatory of Japan, Mitaka, Tokyo 181-8588, Japan
}%
\affiliation{%
$^{4}$
Department of Astronomy, Graduate School of Science, The University of Tokyo,
Hongo 7-3-1, Bunkyo-ku, Tokyo 113-0033, Japan
}%
\affiliation{%
$^{5}$Center for Astrophysics,
Department of Physics, University of Notre Dame, Notre Dame, Indiana 46556, USA.
}%

\date{\today}

\begin{abstract}
We have made an analysis of limits on the neutrino mass based upon the formation of large-scale structure in the presence of a primordial magnetic field. We find that a new upper  bound on the neutrino mass is possible based upon fits to the cosmic microwave background and matter power spectrum when the existing independent constraints on the matter density fluctuation parameter $\sigma_8$ and the primordial magnetic field are taken into account.
\end{abstract}

\maketitle
 \section{Introduction}
The evolution of the cosmic matter density field and large-scale structure (LSS) can be  affected by a finite  neutrino mass \cite{Lesgourgues:2006nd}. 
For example, when the velocities of finite-mass neutrinos are large, the growth of density fluctuations will be impeded \cite{1980PhRvL..45.1980B} on the scale of neutrino free-streaming. 
Cosmological constraints on the neutrino mass are thus of considerable interest. 
Recently an upper limit to the  neutrino mass of 
order $< 0.1-1$ eV 
 has been deduced from a combination of tritium beta-decay endpoint experiments \cite{2008RPPh...71h6201O} and cosmological observations \cite{DeBernardis:2008qq,Ichiki:2008rh,Komatsu:2008hk,Ichiki:2008ye}.

At the same time, 
magnetic fields have been observed 
\cite{Kronberg:1992pp,Wolfe:1992ab,Clarke:2000bz,Xu:2005rb}
in clusters of galaxies  with a strength of $0.1-1.0~\mu$ G. 
One possible explanation for such 
magnetic fields in galactic clusters is the existence of a  primordial magnetic field (PMF) of order  1 nG  whose field lines collapsed as the cluster formed.  
A variety of mechanisms for the origin and amplification of a PMF on the scale of galaxy clusters havs been proposed \cite{Bamba:2004cu,Takahashi:2005nd,ichiki:2006sc} and the associated PMF could have influenced a variety of phenomena in the early Universe
\cite{Grasso:2000wj} such as the cosmic microwave background (CMB)
\cite{
Subramanian:1998fn,
Jedamzik:1999bm,
Mack:2001gc,
Subramanian:2002nh,
Yamazaki:2004vq,
Lewis:2004ef,
Kosowsky:2004zh,
Yamazaki:2005yd,
Challinor:2005ye,
Dolgov:2005ti,
Gopal:2005sg,
Kahniashvili:2005xe,
Kahniashvili:2006hy,
Yamazaki:2006bq,
Yamazaki:2006ah,
Giovannini:2006kc,
Yamazaki:2007oc,
Kojima:2008rf,
2008PhRvD..78f3012K,
Giovannini:2008aa,
Kahniashvili:2008hx,
Yamazaki:2009na,
Kojima:2009ms,
Shaw:2009nf}, 
and the formation of LSS
\cite{Tashiro:2005hc,Tashiro:2005ua,Yamazaki:2006mi,Sethi:2008eq,Yamazaki:2008bb,2008nuco.confE.239Y}.
Of particular relevance to the present work is the fact that the effects of a finite neutrino mass on the matter density fluctuations are degenerate with the effects of a PMF.
Therefore, these two effects must be constrained together when using the fluctuations of the matter density to find limits on  the  neutrino mass and the PMF parameters.

In this regard, the alternative normalization parameter $\sigma_8$ is of particular interest as a measure of large-scale structure effects.  It is defined \cite{Peebles:1980booka} as the root mean square of the matter density fluctuations in a comoving sphere of radius $8h^{-1}$ Mpc.  It is determined by a weighted integral of the matter power spectrum. 
Observations which determine $\sigma_8$ provide information about the physical processes affecting the evolution of density-field fluctuations and the formation of structure on  cosmological scales. 

The detailed mechanisms by which a PMF can affect the density field fluctuations on cosmological scales has been described in Refs.\cite{Yamazaki:2006mi,Yamazaki:2008bb}.  
Of course, $\sigma_8$ is also affected by the presence of a PMF. 
Indeed, in \cite{2008nuco.confE.239Y} we demonstrated for the first time that there was a degeneracy between the parameters of the PMF and the neutrino mass.

In this article, we expand on our previous work \cite{2008nuco.confE.239Y} by also considering the influence on the matter contributions \cite{Kojima:2009ms,Shaw:2009nf}.
By considering the effects of a PMF and finite-mass neutrinos on $\sigma_8$ and comparing theoretically deduced values for $\sigma_8$ with the observed range, we obtain not only insight into the underlying physical processes of density field fluctuations in the presence of a PMF, but also show that the constraint on the sum of neutrino masses from $\sigma_8$ changes when the effects of a PMF are included.
 \section{The Model}
To  evolve the  primary density perturbations  in the presence of a PMF we begin with  magnetized isocurvature initial conditions for the fluids. 
For the present purposes we also fix the cosmological parameters to those of
the best-fit flat $\Lambda$CDM model as given in the WMAP 5 yr analysis in Ref.~\cite{Dunkley:2008ie}, i.e., 
$\Omega_b=0.0462$, $\Omega_c=0.233$, $n_S=0.96$, $h = 0.701$ and $\tau_c=0.084$,
where
$\Omega_b$ and $\Omega_c$ are the baryon and cold dark matter densities in units of the critical density, $n_S$ is the spectral index of the primordial scalar fluctuations, $h$ is the Hubble parameter in units of km s$^{-1}$Mpc$^{-1}$,  and $\tau_c$ is the optical depth for Compton scattering. 

The ionized baryons and electrons are influenced by Lorentz forces when a  PMF is present.
Photons are then indirectly affected by the PMF through Thomson scattering before the epoch of photon last scattering.
We assume that the PMF was generated  earlier during the radiation-dominated epoch. We use a Friedmann-Robertson-Walker (FRW) background cosmology for the linear perturbations. For the time evolution we treat the energy density of the PMF as a first-order perturbation and a stiff source.  Therefore,  all back reactions from the fluid can be discarded.
We also assume that the conductivity of the primordial plasma is very large and that the electric field is negligible, i.e.~$E\sim 0$. 
This ''frozen-in'' condition is  a very good approximation \cite{Mack:2001gc}.
On very large scales the time evolution of the PMF can be decoupled from its spatial dependence, i.e., $\mathbf{B}(\eta,\mathbf{x}) = \mathbf{B_0}(\mathbf{x})/a^2$.
This leads to the following simplified electromagnetic energy-momentum tensor:
\begin{eqnarray}
{T^{00}}_{[\mathrm{EM}]}(\mathbf{x},\tau)=\frac{B(\mathbf{x})^2}{8\pi a^6} \label{eq_MST_00}~~, \\
{T^{i0}}_{[\mathrm{EM}]}(\mathbf{x},\tau)={T^{0k}}_{[\mathrm{EM}]}(\mathbf{x},\tau)=0 \label{eq_MST_0s} ~~,\\
-{T^{ik}}_{[\mathrm{EM}]}(\mathbf{x},\tau)=\sigma^{ik}_\mathrm{B}=
\frac{1}{8\pi a^6}\left\{
	2B^i(\mathbf{x}) B^k(\mathbf{x}) -
\delta^{ik}B(\mathbf{x})^2
\right\}~~,
\label{eq_MST_ss}
\end{eqnarray}
where we use natural units $c=\hbar=1$.  
For a PMF that is statistically homogeneous, isotropic and random, the fluctuation power spectrum can be parametrized as a power-law $P(k) \propto k^{n_\mathrm{B}} $ \cite{Mack:2001gc,Kahniashvili:2006hy} where $n_\mathrm{B}$ is the spectral index. 
A two-point correlation function for the PMF can be defined \cite{Mack:2001gc} by
\begin{eqnarray}
\left\langle B^{i}(\mbi{k}) {B^{j}}^*(\mbi{k}')\right\rangle 
	&=&	\frac{(2\pi)^{n_B+8}}{2k_\lambda^{n+3}}
		\frac{B^2_{\lambda}}{\Gamma\left(\frac{n_B+3}{2}\right)}
		k^{n_B}P^{ij}(k)\delta(\mbi{k}-\mbi{k}'), 
		\ \ k < k_C~~,
		\label{two_point1} 
\end{eqnarray}
where $B_\lambda=|\mathbf{B}_\lambda|$ is the strength of the magnetic comoving mean-field derived by smoothing over a Gaussian sphere of comoving radius $\lambda$
and $k_\lambda = 2\pi/\lambda$ 
 (with $\lambda=1$ Mpc in this paper), and  $P^{i j}(k) \equiv \delta^{ij}- {k{}^{i}k{}^{j}}/{k{}^2}.$
There is a natural cutoff wave number $k_C$ in the magnetic power spectrum 
which is dependent upon the damping scale of the PMF by radiative viscosity. 
It is defined in Refs. \cite{Jedamzik:1996wp,Subramanian:1997gi,Banerjee:2004df}.
Hereafter, we work in k-space and denote all quantities by their Fourier transform convention
$F(\mbi{k})=\int d^3 x \exp (i\mbi{k} \cdot \mbi{x})F(\mbi{x}). $

We evaluate the PMF source power spectrum using the numerical methods described in 
Refs.~\cite{Yamazaki:2006mi,Yamazaki:2007oc}.  
Using this, we can quantitatively evaluate the time evolution of the cut off scale and thereby more reliably calculate the effects of a PMF.

We adopt the model of Ref.\cite{Kojima:2008rf,Kojima:2009ms} for estimating the effects of finite-mass neutrinos and the PMF on fluctuations of the matter density fields in the early Universe. We also utilize adiabatic initial conditions for the matter contributions as in Ref.\cite{Kojima:2009ms,Shaw:2009nf}. This leads to stable numerical calculations of the curvature perturbation of the scalar mode for all scales and times.  This is an improvement over previous numerical estimations for which the scalar curvature perturbations were too small to stabilize numerical calculations for large scales and early times. Thus, we can obtain consistent results for all scales and times of interest in the present work.
 \section{Results and Discussions}
Details of the effects of a PMF on the cosmological density field fluctuations can be found in Ref.~\cite{Yamazaki:2006mi}.  
Here we briefly review these effects and show that a relation exists between $\sigma_8$ and the PMF parameters.
 The observational constraints \cite{Cole:2005sx,Tegmark:2006az,Rozo:2007yt,Ross:2008ze} 
  on  $\sigma_8$,  therefore,  lead to  a strong constraint on the PMF parameters and the neutrino mass. 
Since the parameters of the PMF have a strong degeneracy, the existence of a Bayesian prior constraint on $\sigma_8$ can be used to effectively constrain the PMF. Also, since $\sigma_8$ is constrained by diverse observational data on linear cosmological scales, we can obtain a reliable prior 
for use in determining the probability distributions for the parameters of the PMF from CMB observations.

The PMF effects dominate the matter power spectrum for wave numbers $k >  0.1 h$ Mpc$^{-1}$\cite{Yamazaki:2006mi}(Fig.1).  This is because the PMF energy density fluctuations depend only on the scale factor $a$ and can survive below the Silk damping scale. Therefore, the PMF continues to be a source for the fluctuations via the Lorentz force even below the Silk damping scale.
In the case of no correlation between the PMF and the matter density fluctuations, the matter power spectrum is increased by the PMF independently of whether the PMF is dominated by the pressure or the magnetic tension\cite{Yamazaki:2006mi}.

A PMF affects the power spectrum function $P(k)$ and the matter density fluctuation $\delta$ differently. 
The total density fluctuation $\delta$ can be smaller or larger
depending on whether the effect of the PMF is mainly due to  its pressure or tension. On the other hand, the power spectrum function $P(k)$ always increases
in the case that the PMF does not correlate with the primordial density fluctuations.  This is  because $P(k)\propto \delta^2$ and is not affected by the sign of $\delta$.
 \subsection{PMF parameters and $\sigma_8$}
The alternative normalization parameter $\sigma_8$ can be determined from a  study of the physical processes of density field fluctuations on cosmological scales within the linear regime.
Recently $\sigma_8$ has been constrained in this way by observations \cite{Cole:2005sx,Tegmark:2006az,Rozo:2007yt,Ross:2008ze} to be in the range  $0.7 < \sigma_8 < 0.9$.
From this we can obtain strong constraints on  the PMF parameters by numerically calculating $\sigma_8$ under the influence of a PMF.

We have shown \cite{Yamazaki:2006bq,Yamazaki:2006mi,Yamazaki:2007oc} that the possible discrepancy between theoretical estimates and observational temperature fluctuations of the CMB for higher multipoles ($\ell > 1000 $) could be attributed to  a PMF of comoving strength 1.0 nG$< B_\lambda<$2.0nG. Since other constraints on the amplitude of the PMF imply $B_\lambda \le$10 nG\cite{Subramanian:1998fn,Jedamzik:1999bm,Mack:2001gc,Lewis:2004ef,Kahniashvili:2008hx,Schleicher:2008dt}, our allowed range is consistent with previous works.

The value of  $\sigma_8$ derived by a CMB analysis including such a field strength for the PMF is also $0.78 < \sigma_8 <  0.84$. 
As noted above, in this work we consider $\sigma_8$ constraints from both fits to the CMB and from large-scale structure. In this case, we adopt a prior for $\sigma_8$ in the range of  0.75$< \sigma_8 < $0.85 as a reasonable average of constraints from observed structure and fits to the CMB.
\subsection{Constraints on the neutrino mass and PMF parameters from $\sigma_8$}
Figure \ref{fig2} shows the constraints on the PMF parameter $B_\lambda$ and the sum of neutrino masses $\Sigma m_\nu$  for various fixed values of  $n_\mathrm{B}$ and ranges of $\sigma_8$ as labeled.
 Since the PMF power spectrum depends on $n_\mathrm{B}$, PMF effects on
the density fluctuations for small scales decrease with lower values for $n_\mathrm{B}$.
 Based upon previous work \cite{DeBernardis:2008qq,Ichiki:2008rh,Komatsu:2008hk,Ichiki:2008ye} the upper limit on the total neutrino mass is expected to be of order $< 0.1-1$ eV.
For such masses the neutrino velocities can be quite  large while the mass density
of neutrinos is still a significant fraction of the total dark matter density.   Therefore,
the growth of density fluctuations on the free-streaming scale of neutrinos  will  be hindered.

\begin{figure*}[h]
\includegraphics[width=1.0\textwidth]{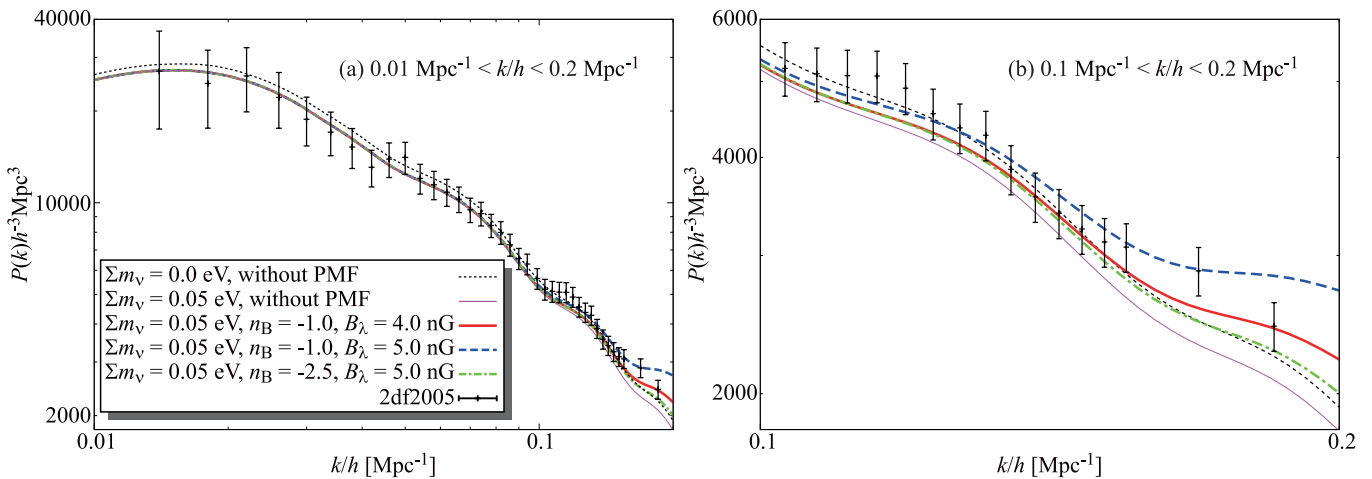}
\caption{\label{fig1} 
Comparison of the effects of a PMF and finite-mass neutrinos on the matter power spectrum.
The left panel is for the full power spectrum on scales $0.01 \mathrm{Mpc}^{-1} < k/h < 0.2 \mathrm{Mpc}^{-1}$ and various parameters as labeled.
The right panel shows the expanded spectrum in the range of $0.1 \mathrm{Mpc}^{-1} < k/h < 0.2 \mathrm{Mpc}^{-1}$.
Curves and dots with error bars in all panels are as indicated in the legend box. 
In order to better illustrate the effects of a PMF and a finite neutrino mass on the matter power spectrum, we have utilized a  larger value for the  PMF amplitude $B_\lambda$.
This figure shows that the matter power spectrum for $k/h > 0.1 \mathrm{Mpc}^{-1}$ is affected strongly by a PMF.  The magnitude of the effect depends on both  the PMF amplitude $B_\lambda$ and the power spectral index $n_\mathrm{B}$. On the other hand, the mass of the neutrinos decreases the total amplitude of the matter power spectrum. Also, the effects of a neutrino mass  for smaller scales (large wave number $k$) are greater than for larger scales (small $k$).
}
\end{figure*}

\begin{figure*}[h]
\includegraphics[width=1.0\textwidth]{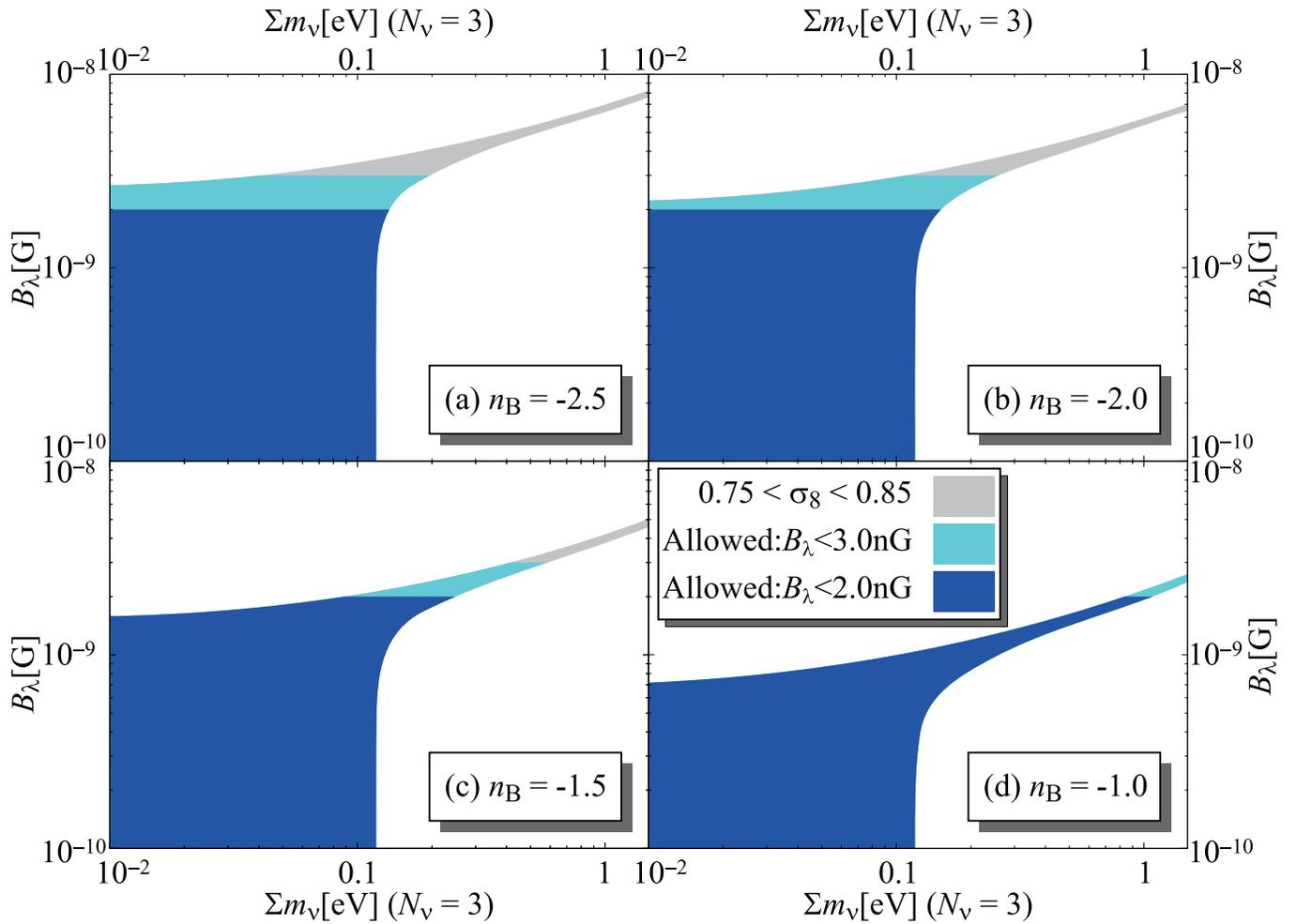}
\caption{\label{fig2} 
Excluded and allowed regions in the parameter plane of PMF amplitude $B_\lambda$ vs mass of neutrinos $\sum_{N_\nu = 3} m_\nu $.
Shaded regions in all panels are allowed by the indicated  ranges for $\sigma_8$ and $B_\lambda$.
Various panels are for the indicated values of the PMF power-law spectral index $n_\mathrm{B}$. 
}
\end{figure*}

Thus, neutrinos decrease the matter density fluctuations, while a PMF increases the matter density fluctuations (cf.~Figure \ref{fig1}). 
Furthermore, for $n_\mathrm{B}$ within the ranges constrained by previous work \cite{Yamazaki:2004vq,Yamazaki:2006bq}, a PMF with $B_{\lambda} ~^> _\sim 1$ nG affects the matter density fluctuations (see Fig.~\ref{fig2}).
When the mass of the neutrinos is constrained from matter density fluctuations in the presence of a PMF, 
the deduced upper limit is heavier with a PMF than  without. 

The expected parameters of the PMF from the combined analysis of the CMB and observed magnetic fields in galactic clusters is 
$B_\lambda<$2.0nG$(1 \sigma)$ and $<$3.0nG$(2 \sigma)$
\cite{Yamazaki:2004vq,Yamazaki:2006bq,Yamazaki:2009na}, while the expected value of $\sigma_8$ based upon observations is 0.75$< \sigma_8 < $0.85 as noted above.
For this range of $\sigma_8$, the sum of the neutrino masses is constrained to be 
$\sum_{N_\nu} m_\nu < 0.8$ eV (from the 1$\sigma$ constraint on  $B_\lambda$) and 
$< 2.2$eV (from the 2$\sigma$ constraint on $B_\lambda$)  for $N_\nu = 3$.  
This is a larger upper limit than that deduced previously because the effect of the PMF cancels the effect of neutrinos on the density fluctuations \cite{DeBernardis:2008qq,Ichiki:2008rh,Komatsu:2008hk,Ichiki:2008ye}.
\subsection{Future Work}
Komatsu et al. (2008) noted that the effect of a finite neutrino mass is very small for the detection range of the WMAP, e.g. $\ell < 1000$\cite{Komatsu:2008hk}. 
Therefore, as a first step toward constraining the neutrino mass together with the PMF we can use the WMAP CMB power spectrum for $\ell < 1000$ to obtain values for the primary cosmological parameters without concern for a degeneracy between these parameters and the neutrino mass.
One cannot constrain the mass of neutrinos very well from such CMB data alone.
Hence, one must use other observational data which are independent of the CMB analysis to constrain the neutrino mass.
For this purpose, $\sigma_8$ is very useful since it is determined from independent observations of LSS in the regime of linear perturbation theory. 
Since the main purpose of the present work is to examine how the upper limit of the neutrino mass determined from $\sigma_8$ changes when a PMF is included, we have focused on a theoretical analyses of effects of both the PMF and the neutrino mass on $\sigma_8$ and the matter power spectrum. 
However, theoretical estimates of $\sigma_8$ depend upon the cosmological model employed and 
$\sigma_8$ has a significant degeneracy with 
$\Omega_m = \Omega_b + \Omega_\mathrm{CDM}$, 
$n_\mathrm{S}$, and 
$A_\mathrm{S}$, even if such primary parameters are well constrained by the CMB data, e.g. WMAP\cite{Komatsu:2008hk}. 
Furthermore, there is the degeneracy between the mass of neutrinos and $\Omega_m$
\cite{Elgaroy:2002bi}. 
Therefore, one should ultimately consider the degeneracy between the PMF and other cosmological parameters including the mass of neutrinos.
In subsequent work we will carry out a full-constraint on the mass of neutrinos using a Markov chain Monte Carlo (MCMC) method applied to both the CMB and the LSS data. 
In that effort we will need to consider the model dependence and degeneracy between cosmological parameters including the neutrino mass and PMF parameters.
\section{Summary}
Previous numerical estimates of curvature perturbations without matter contributions were too small at early times to stabilize on large scales. 
In the present work we have stabilized the time evolution of curvature perturbations in the presence of a PMF while considering the matter contributions. 
We find that a PMF significantly affects the constraint on the sum of neutrino masses. 
We confirm that the upper limit on the neutrino mass from $\sigma_8$ in the presence of a PMF is heavier than without a PMF even if we consider the matter contributions. 
We also have shown that the prior limited range on the sum of neutrino masses and PMF parameters is within the expected range for $\sigma_8$ from observations of the LSS.
Since the sum of the neutrino masses has a degeneracy with some of the cosmological parameters, e.g. $\Omega_m$, one should ultimately utilize a more complete MCMC method with the CMB and LSS data to resolve such degeneracies and obtain a more accurate constraint on the sum of the neutrino masses.
Constraints on the PMF parameters and $\sigma_8$ are soon to be improved by future cosmological observational programs such as Quiet, Planck, and the Sloan Digital Sky Survey (SDSS). 
In principle, by applying our method to such data it will be possible to obtain not only the upper but also the lower limits to the neutrino mass from cosmology in the presence of a PMF.

\begin{acknowledgments}
This work has been supported in part by Grants-in-Aid for Scientific
Research (20105004, 20244035, 21740177) of the Ministry of Education, Culture, Sports,
Science and Technology of Japan.  This
work is also supported by the JSPS Core-to-Core Program, International
Research Network for Exotic Femto Systems (EFES).  Work at UND is supported in part by the US Department of Energy under research Grant No. DE-FG02-95-ER40934.
\end{acknowledgments}

\bibliographystyle{apsrev}

\end{document}